\magnification1200
\rightline{KCL-MTH-03-09}
\rightline{hep-th//0307098}

\vskip .5cm
\centerline
{\bf   $E_{11}$, SL(32) and  Central Charges}
\vskip 1cm
\centerline{  Peter West }
\vskip .5cm
\centerline{Department of Mathematics}
\centerline{King's College, London, UK}

\leftline{\sl Abstract}
\noindent 
We  show that  the $E_{11}$ representation that contains the space-time
translation generators  also contains the rank two and five totally
anti-symmetric representations of $A_{10}$.  By studying the behaviour of
these latter $A_{10}$ representations  under SL(32), which we argue is
contained in   the Cartan involution invariant sub-algebra of
$E_{11}$,  we find that the rank two and five totally
anti-symmetric representations must be identified with
the central charges of the eleven dimensional supersymmetry algebra.

\vskip .5cm

\vfill

\vskip 1cm
email:  pwest@mth.kcl.ac.uk

\eject

\medskip
{\bf {1 Introduction}}
\medskip
It has been shown that the entire bosonic sector of eleven dimensional
supergravity,  can  be formulated as non-linear
realisation [1]. The algebra, denoted $G_{11}$, used for this  non-linear
realisation was not a Kac-Moody algebra. However, it has been suggested
[2] that  this theory can be reformulated, or extended, in such a way
that it can be described as  a non-linear realisation based on a Kac-Moody
algebra.  Assuming this to be the case,     it was   shown [2] that
this Kac-Moody symmetry should contain  very extended
$E_8$, i.e.  $E_{11}$. Substantial fragments of
this symmetry, as well as other evidence for it, has been presented
[2,3,4]. Analogous results were also found for IIA [2] and IIB
supergravity [5] where the corresponding Kac-Moody algebras were also
found to be
$E_{11}$ in each case. An account of  work on
symmetries in dimensionally reduced supergravity and string theories is
given in reference [15]. 
\par
The algebra $G_{11}$ contained the usual translation generators  whose
role was  to introduce space-time into the the theory, however, these
generators did not play any role in subsequent discussions on $E_{11}$.
In this paper we will  explain how the translation 
generators  can
be incorporated into a non-linear realisation based on $E_{11}$. 
The translation generators carry  one lower index and transform in the 
corresponding    representation of
$A_{10}$.  Since $E_{11}$ contains
$A_{10}$ we will enlarge this $A_{10}$ representation to
 one belonging to  
$E_{11}$.   We will find that the translations occur in this $E_{11}$ 
representation together with a second rank and fifth rank anti-symmetric
tensor representations of $A_{10}$ as well as an infinite number of other
tensors.  
\par
The gravity sector of eleven
dimensional supergravity arises as the non-linear realisation of the 
IGL(11) sub-algebra of $G_{11}$, the I being the translations.  
Taking the  non-linear realisation of IGL(11) with the  Lorentz algebra 
SO(1,10) as the local sub-algebra does not uniquely lead to gravity. 
However,  if one takes the 
simultaneous non-linear realisation of IGL(11) and the conformal
group SO(2,11)  one finds    to Einstein's theory
of gravity [6,1] essentially uniquely.    In reference [1] the non-linear
realisation based on
$G_{11}$ used for eleven dimensional supergravity was  combined with a 
simultaneous non-linear realisation of  the conformal
algebra and this   did lead to the unique
bosonic field equations of  eleven dimensional supergravity, up to one
undetermined constant [1]. 
Subsequent papers did not
address the question of how this conformal algebra combined with
$E_{11}$. However, it was argued in reference [1] that the presence of
the  fermionic extension  of bosonic sector of eleven dimensional
supergravity considered above implied that the full theory would possess 
an Osp(1/64).  This algebra contains  GL(32) which
rotates the spinor index on the supercharges and so this must also be a
symmetry of M theory.  In fact, the algebra GL(32) had previously been
proposed as a symmetry of M theory as part of it occurred as a symmetry
of the fivebrane equations of motion [8]. It was explained [8] that the 
SL(32) symmetry was a brane rotating symmetry and was the natural
generalisation of the local spin or Lorentz algebra when branes were
present. 
\par
The local sub-algebra used in 
formulating the $E_{11}$ non-linear realisation was taken to be the one
which is invariant under the Cartan involution. 
 We will argue that SL(32) is indeed part of this local sub-algebra  and
we calculate the transformations of  the  second  and fifth rank
anti-symmetric tensor representations mentioned above  under SL(32). 
We will find that these objects transform in such a way that they 
should be    identified   with central charges  of the eleven dimensional
supersymmetry algebra.

\medskip
{\bf 2 {$E_{11}$ at Low Levels}}
\medskip
We invite the reader to draw the   Dynkin diagram of $E_{11}$ by drawing 
 ten dots in  a horizontal line labeled from one to ten from left to
right and connected     by a single line. Then place another dot,
labeled eleven,   above the third node (labeled eight)  from the
right.  We consider $E_{11}$ as a member of the  class of Kac-Moody
algebras discussed in section three of reference [7], namely an algebra
whose Dynkin diagram  possess at least  one   node such that deleting it
leads to a finite dimensional semi-simple Lie algebra. 
If we delete  node eleven in the Dynkin diagram of  $E_{11}$, the
remaining algebra is
$A_{10}$.  The preferred simple root is 
$\alpha_{11}$ and the simple roots of $A_{10}$ are   $\alpha_i,
i=1,\ldots , 10$. We may write  [7]   
$$\alpha_{11}=-\lambda_8 +x    
\eqno(2.1)$$
where  $x$ is a vector in a space
orthogonal to the roots of $A_{10}$ and $\lambda_i$ are the fundamental
weight vectors  of
$A_{10}$. The simple roots have length squared two and so 
 $x^2=2-\lambda_8^2=-{2\over 11}$. 
\par
A  root $\alpha$ of  $E_{11}$  can be
written as
$$\alpha= l\alpha_{11} +\sum_im_i\alpha_i
=lx-l\lambda_8+\sum_{jk} A^f_{jk}\lambda_k
\eqno(2.2)$$
where $A^f_{jk}$ is the Cartan matrix of $A_{10}$.
We define the level, denoted $l$, [4] of the roots of $E_{11}$ to be the
number of times the root $\alpha_{11}$ occurs in its decomposition into
simple roots given in the equation above. The generators of
$E_{11}$ can also be classified according 
to their level which is just the
level of the
 root associated with the generator. 
\par 
The $E_{11}$ algebra contains the  generators  $K^a{}_b$ at level 0
and the generators 
$$ R^{a_1a_2a_3}, R^{a_1a_2\dots a_6},R^{a_1a_2\ldots a_8,b}
\eqno(2.3)$$
 at levels zero, 1, 2 and 3
repectively [2,4], as well as  the generators 
$$ R_{a_1a_2a_3}, R_{a_1a_2\dots a_6},R_{a_1a_2\ldots a_8,b}
\eqno(2.4)$$
at levels -1, -2,-3.    
The generators of $E_{11}$ at higher levels are listed in references
[4,16]. 
\par 
The corresponding Borel sub-algebra up to, and including,
level 3 obeys the commutation relations [2]
$$
[K^a{}_b,K^c{}_d]=\delta _b^c K^a{}_d - \delta _d^a K^c{}_b,  
\eqno(2.5)$$
$$  [K^a{}_b, R^{c_1\ldots c_6}]= 
\delta _b^{c_1}R^{ac_2\ldots c_6}+\dots, \  
 [K^a{}_b, R^{c_1\ldots c_3}]= \delta _b^{c_1}R^{a c_2 c_3}+\dots,
\eqno(2.6)$$
$$ [ K^a{}_b,  R^{c_1\ldots c_8, d} ]= 
(\delta ^{c_1}_b R^{a c_2\ldots c_8, d} +\cdots) + \delta _b^d
R^{c_1\ldots c_8, a} .
\eqno(2.7)$$
$$[ R^{c_1\ldots c_3}, R^{c_4\ldots c_6}]= 2 R^{c_1\ldots c_6},\ 
\ \ 
[R^{a_1\ldots a_6}, R^{b_1\ldots b_3}]
= 3  R^{a_1\ldots a_6 [b_1 b_2,b_3]}, 
\eqno(2.8)$$
where $+\ldots $ means the appropriate anti-symmetrisation. 
The level 0 and negative level generators obey the relations 
$$[K^a{}_b, R_{c_1\ldots c_3}]= -\delta ^a_{c_1}R_{b c_2
c_3}-\dots,\ [K^a{}_b, R_{c_1\ldots c_6}]=  -\delta ^a_{c_1}R_{bc_2\ldots
c_6}-\dots,
\eqno(2.9)$$
$$ [ K^a{}_b,  R_{c_1\ldots c_8, d} ]= 
-(\delta ^a_{c_1} R_{b c_2\ldots c_8, d} +\cdots) - \delta ^a_d
R_{c_1\ldots c_8, b} .
\eqno(2.10)$$
$$[ R_{c_1\ldots c_3}, R_{c_4\ldots c_6}]= 2 R_{c_1\ldots c_6},\ 
\ \ 
[R_{a_1\ldots a_6}, R_{b_1\ldots b_3}]
= 3  R_{a_1\ldots a_6 [b_1 b_2,b_3]}, 
\eqno(2.11)$$
Finally, the commutation relations between the positive and negative
generators of  up  to level three  are given by 

$$[ R^{a_1\ldots a_3}, R_{b_1\ldots b_3}]= 36 \delta^{[a_1a_2}_{[b_1b_2}
K^{a_3]}{}_{b_3]}-4\delta^{a_1a_2 a_3}_{b_1b_2 b_3} D,\  
[ R_{b_1\ldots b_3}, R^{a_1\ldots a_6}]= -{6!\over 3!}
\delta^{[a_1a_2a_3}_{b_1b_2b_3}R^{a_4a_5a_6]}
\eqno(2.12)$$
where $D=\sum_b K^b{}_b$, $\delta^{a_1a_2}_{b_1b_2}=
{1\over
2}(\delta^{a_1}_{b_1}\delta^{a_2}_{b_2}-
\delta^{a_2}_{b_1}\delta^{a_1}_{b_2})=
\delta^{[a_1}_{b_1}\delta^{a_2]}_{b_2}$ with similar formulae when 
more indices are involved. 
\par
The above commutators can be deduced, using the Serre relations and   from
the identification of the Chevalley generators of $E_{11}$ which are given
by [2] 
$$E_a=K^a{}_{a+1}, a =1, \ldots 10, \ 
E_{11}= R^{91011}; \ \ F_a=K^{a+1}{}_{a},  a =1, \ldots 10,\ 
F_{11}= R_{91011}
\eqno(2.13)$$
$$ H_a= K^a{}_a-K^{a+1}{}_{a+1}, a=1,\ldots ,10, \ 
H_{11}={2\over 3}(K^9{}_9 +K^{10}{}_{10}+K^{11}{}_{11})
-{1\over 3}(K^1{}_1+\ldots +K^8{}_8). 
\eqno(2.14)$$
\par
The sub-algebra which is invariant under the Cartan involution, namely 
 $$E_a\to -F_a, F_a\to -E_a, H_a\to -H_a, 
\eqno(2.15)$$ plays an important part in the
non-linear realisation of reference [2] as it is taken to be the local
sub-algebra. As such, its generators do not lead  to fields in 
the non-linear realisation.  The Cartan involution is a
linear operator and acts  on the generators as 
$$K^a{}_b\to -K^b{}_a,\ R^{a_1a_2a_3}\to -R_{a_1a_2a_3},
\ R^{a_1\ldots a_6}\to R_{a_1\ldots a_6},\ R^{a_1\ldots a_8,b}\to
-R_{a_1\ldots a_8,b}
\eqno(2.16)$$
The sub-algebra invariant under the Cartan involution  is generated by  
$E_a-F_a$ and at low levels it includes the generators  
$$
J_{ab}=K^c{}_b\eta_{ac}-K^c{}_a\eta_{bc},\ 
S_{a_1a_2a_3}=R^{b_1b_2b_3}\eta_{b_1a_1}\eta_{b_2a_2}\eta_{b_3a_3}
-R_{a_1a_2a_3},
\eqno(2.17)$$
$$
S_{a_1\ldots a_6}=R^{b_1\ldots b_6}\eta_{b_1a_1}\ldots \eta_{b_6a_6}
+R_{a_1\ldots a_6}
\eqno(2.18)$$
$$
S_{a_1\ldots a_8,c}=R^{b_1\ldots b_8,b}\eta_{b_1a_1}\ldots \eta_{b_8a_8}
\eta_{b c}-R_{a_1\ldots a_8,c} .
\eqno(2.19)$$
The generators $J_{ab}$ are those of the Lorentz algebra SO(1,10) and
their commutators with   the other generators just express the fact that
they belong to a   representation of the Lorentz algebra. 
 The $S_{a_1a_2a_3}$  and $S_{a_1\ldots a_6}$
generators obey the commutators
$$
[ S^{a_1a_2a_3}, S_{b_1b_2b_3}]=-36\delta^{[a_1a_2}_{[b_1b_2}
J^{a_3]}{}_{b_3]}+2S^{a_1a_2a_3}{}_{b_1b_2b_3}
\eqno(2.20)$$
$$
[ S_{a_1a_2a_3}, S^{b_1\ldots b_6}]=-2.{6!\over
3}\delta^{[b_1b_2b_3}_{a_1a_2a_3} S^{b_4b_5b_6]}-3S^{b_1\ldots
b_6}{}_{[a_1a_2,a_3]}
\eqno(2.21)$$
\medskip
{\bf {3 Translations and an $E_{11}$ Representation}}
\medskip
 The space-time translation generators 
carry  a  single lower index and transform in  the
corresponding $A_{10}$ representation. 
 This is  equivalent to a tensor with ten upper
anti-symmetrised indices which, in our conventions,  is the
representation   with weight
$\lambda_1$,   or Dynkin index $p_1=1$, all other $p_i$'s vanishing. 
We wish to consider the representation of $E_{11}$ 
which contain the translation generators. 
The fundamental weights  of $E_{11}$ are given by [7]
$$ l_i=\lambda_i+\lambda_8.\lambda_i{x\over x^2},\ i=1,\ldots ,10\ \ 
l_{11}={x\over x^2}
\eqno(3.1)$$
The $E_{11}$
representation with highest weight $l_1=\lambda_1-{3\over 2}x$ 
obviously  contains the states in the $A_{10}$ representation
$\lambda_1$. The highest weight state 
$|l_1>$ can be thought of as being at level
$-{3\over 2}$. It is straightforward to construct the root string 
associated
with the action of the simple negative roots  
$F_a$ on this highest weight state. One finds 
$$
l_1,l_1-\alpha_1,l_1-\alpha_1-\alpha_2, \ldots , l_1-\alpha_1-\ldots
\alpha_8-\alpha_{11},
\ldots.
\eqno(3.2)$$ 
The last weight  written explicitly is the first one in the string where
$\alpha_{11}$  enters and it corresponds to the appearance of a new
$A_{10}$ representation in the string.  In fact,  
$$l_1-\alpha_1-\ldots -\alpha_8-\alpha_{11}=\lambda_9-{5\over 2}x,
\eqno(3.3)$$
which contains the highest weight for  the
$A_{10}$  representation whose only non-vanishing  Dynkin index is 
$p_9=1$ or a second rank anti-symmetric tensor. Continuing in this way we
find that the representation $l_1$ contains the following $A_{10}$
representations 
$$  p_{1}=1, \ (-{3\over 2});p_{9}=1, \ (-{5\over 2}); p_{6}=1, \
(-{7\over 2});, 
  p_{4}=1, p_{10}=1, \ (-{9\over 2});\ {\rm and }\ 
p_3=1, \ (-{9\over 2});\ldots , 
\eqno(3.4)$$
all other $p_i$'s vanishing. The number in brackets is the corresponding
level. 
\par
Consider any Lie algebra ${ g}$ with a   representation $u(A)$. 
By
definition, the linear operators $u(A)$, for each
element $A\ \in g$, obey the relation
$u(A_1A_2)=u(A_1)u(A_2)$. If the representation is  carried by the states 
$|X_s>$ it defines the matrices $u(A)|X_s>=(c(A))_s{}^t|\chi_t>$. 
Clearly, $c(A_1A_2)=c(A_2) c(A_1)$ and so a matrix representation of $g$
is defined by $d(A)=c(A^\ddagger)$  where $\ddagger$ is any operation that
inverts the order of the factors in the Lie algebra. 
The relevant operation for us is to take $A^\ddagger=I^cI^I(A)$
where $I^c$ is the Cartan involution and $I^I$ is the operators which
inverts the order of operators and $I^I(A)=-A$. The latter operator is
just the operator in the algebra which  corresponds to inversion of
group elements.  
\par
In these circumstance  we can define a  semi-direct product algebra. We
associate with each state in the representation
$|X_s>$ a generator
$X_s$ and  we extend the algebra
$g$ to include the new generators by adopting the commutation relation
$$
[X_s, A]=d(A)_s{}^r X_r, A\in g
\eqno(3.5)$$
This is consistent with the Jacobi identities involving two elements of
$g$ and one $X_s$. The commutator between two elements $X_s$ and $X_r$
must be chosen and  it is consistent to  choose
it to vanish.
\par
Carrying out this procedure for $E_{11}$ and the representation $l_1$ we
introduce the  generators 
$$
P^{a_1\ldots a_{10}}, W^{a_1a_2}, W^{a_1\ldots a_5}, 
W^{a_1\ldots a_7,b},\dots 
\eqno(3.6)$$
at levels ${3\over2},{5\over2},{7\over2},{9\over2}$ respectively. 
it is straightforward to find the generator associated with the root
string of equation (3.2). The entries explicitly given correspond to
$P^{2\ldots 11}$, $P^{13\ldots 11}$ , $P^{124\ldots 11}$ and $R^{91011}$.
We denote the semi-direct product of 
$E_{11}$ with its   $l_1$  representation by $E_{11}\oplus_s L_1$ 
\par 
It is simpler to work with the more familiar
generator
$$P_a={1\over 10!}\epsilon_{b a_1\ldots a_{10}}P^{a_1\ldots a_{10}}
\eqno(3.7)$$
The $A_{10}$ generators act on the translations as 
$$
[K^c{}_b, P^{a_1\ldots a_{10}}]=10\delta^{[a_1}_{b}P^{|c| a_2\ldots
a_{10}]}-{1\over 2}\delta^c_b  P^{a_1\ldots a_{10}}
\eqno(3.8)$$
The last term on the right-hand side of the above equation  is
required as a result of  the relation 
$[ H_{11},  P^{a_1\ldots a_{10}}]=0$.  
The corresponding commutator involving $P_a$ is 
$$
[K^c{}_b, P_a]=-\delta_a^c P_b+{1\over 2}\delta^c_b P_a
\eqno(3.9)$$
\par
The root string of equation (3.2) corresponds to the commutators  
 $$[R^{a_1a_2a_3},P_b]=3\delta_b^{[a_1}W^{a_2a_3]},\ 
[R^{a_1a_2a_3},W^{b_1b_2}]=W^{a_1a_2a_3b_1b_2},
\eqno(3.10)$$
$$
[R^{a_1a_2a_3},W^{b_1\ldots b_5}]=W^{b_1\ldots b_5 [a_1a_2,a_3]}
\eqno(3.11)$$
which also normalise the new generators. 
\par
Using these relationships and the Jacobi identities one deduces that 
$$
[R^{a_1\ldots a_6},P_b]=-3\delta_b^{[a_1}W^{a_2\ldots a_6]}
\eqno(3.12)$$
Using the Jacobi identities and equations (3.10) and (3.11) the
commutators involving  the negative generators 
$R_{abc}$ is found to be given by 
$$[R_{a_1a_2a_3}, P_b]=0,\ [R_{a_1a_2a_3}, W^{b_1b_2}]= 12 \delta
^{b_1b_2}_{[a_1a_2} P_{a_3]}
\eqno(3.13)$$
$$[R_{a_1a_2a_3}, W^{b_1\ldots b_5}]= 120 \delta
^{[b_1b_2b_3}_{a_1a_2a_3} W^{b_4b_5]}
\eqno(3.14)$$
\par
It will be instructive to consider the commutators  of the Cartan
invariant sub-algebra generator $S_{abc}$ of  $E_{11}$ with the
generators associated with the representation $l_1$. Using equations
 (3.10) and (3.11) and  equations (3.13) and (3.14) we find that 
$$[S_{a_1a_2a_3}, P_b]=3\eta_{b[a_1}W_{a_2a_3]},\ 
[S_{a_1a_2a_3}, W^{b_1b_2}]=W_{a_1a_2a_3}{}^{b_1b_2}-12\delta^{b_1b_2}_
{[a_1a_2}P_{a_3]}
\eqno(3.15)$$
$$[S_{a_1a_2a_3},
W^{b_1\ldots
b_5}]=W^{b_1\ldots b_5}{}_{[a_1a_2, a_3]}-120\delta^{[b_1b_2b_3}
_{a_1a_2a_3}W^{b_4b_5]}
\eqno(3.16)$$
\par
It is interesting to examine what adding the generators corresponding to
the $l_1$ representation means in terms of the weight lattice. The weight 
$l_1=-{1\over 2}(l+\bar l)$ in the notation of
reference [7], section five. The root lattice of $E_{11}$ is given by 
[7] 
$$\Lambda_{E_8}\oplus \Pi^{(1,1)}\oplus \{ (n,n),\ n\in \ {\bf Z}\}
\eqno(3.17)$$
Adding the $l_1$ representation corresponds to adding the vector 
$-{1\over 2}(1,-1)$ to the last factor in the above decomposition and so
 one is working on the  full weight lattice. 

\medskip
{\bf {4 GL(32) and   Central Charges }}
\medskip
In this section we review and  expand the results of
references [8], [9] and [1]. The eleven dimensional supersymmetry
algebra is of the form [10] 
$$\{Q_\alpha, Q_\beta\}=Z_{\alpha\beta}=( \gamma^a C^{-1}
P_a+\gamma^{a_1a_2} C^{-1}Z_{a_1a_2}+
\gamma^{a_1\ldots a_5}C^{-1}Z_{a_1\ldots a_5})_{\alpha\beta}
\eqno(4.1)$$
$$ 
[Q_\alpha, Z_{\alpha\beta}]=0,[Z_{\alpha\beta},Z_{\gamma\delta}]=0
\eqno(4.2)$$
This algebra admits  GL(32) as an 
automorphism whose   action is given by 
$$[Q_\alpha, T_\gamma{}^\delta ]= \delta_\alpha{}^\delta Q_\gamma, \ 
[Z_{\alpha\beta},  T_\gamma{}^\delta ]
=  \delta_\alpha^\delta Z_{\gamma\beta}+
\delta_\beta{}^\delta  Z_{\alpha\gamma}
\eqno(4.3)$$
This automorphism was found to play  a role in the symmetries of the
fivebrane dynamics [8], [9] and later was  shown to be a symmetry
of M theory [1]. 
\par 
To
gain a more familiar set of generators we may expand
$T_\gamma{}^\delta$ out in terms of the elements of the enveloping
Clifford algebra
$$
T_\gamma{}^\delta= \sum_p \sum_{a_1\ldots a_p}
{1\over 32 p!}(-1)^{{p(p-1)\over 2}}(\gamma^{a_1\ldots
a_p})_\gamma{}^\delta  T_{a_1\ldots a_p} \ \ {\rm {or}}\ \ 
(\gamma^{a_1\ldots a_p})_\delta{}^\gamma T_\gamma{}^\delta=
 T^{a_1\ldots a_p}.
\eqno(4.4)$$
The summation only goes over $p=0,1,\ldots,4,6$ since we may  
use the relation 
$$\gamma^{a_1\ldots a_p}={1\over q!}(-1)^{{(q-1)p}}(-1)^{{q(q+1)\over 2}}
\epsilon^{a_1\ldots a_p b_1\ldots b_q}\gamma_{b_1\ldots b_q},
\eqno(4.5)$$
for $p+q=11$. We then find that 
$$
[Q_\alpha  ,T^{a_1\ldots a_p}]=(\gamma^{a_1\ldots
a_p})_\alpha{}^\beta Q_\beta 
\eqno(4.6)$$
As such,  the $Q_\alpha$ carry a representation of GL(32) with 
$T^{a_1\ldots a_p}$ represented by $\gamma^{a_1\ldots a_p}$. 
The Lorentz generators are represented in the usual way by 
$J_{ab}=\gamma_{ab}$. 
\par
Using the relation 
$$
[\gamma^{a_1\ldots a_n},\gamma_{b_1\ldots b_m}]=
\sum_{s=0}^{n}{n!m!\over s! (n-s)!(m-s)!}(-1)^{{s(s+1)\over 2}} (-1)^{sn}
$$
$$(1-(-1)^{(nm+s)})
\delta^{[a_1\ldots a_s}_{[b_1\ldots b_s}
\gamma^{a_{s+1}\ldots a_n]}{}_{b_{s+1}\ldots b_m]},
\eqno(4.7)$$
valid for $m>n$, it straightforward to show, for example,  that 
$$
[ T^{a_1a_2a_3}, T_{b_1b_2b_3}]=-36\delta^{[a_1a_2}_{[b_1b_2}
J^{a_3]}{}_{b_3]}+2T^{a_1a_2a_3}{}_{b_1b_2b_3}
\eqno(4.8)$$
$$
[ T^{a_1a_2a_3}, T_{b_1\ldots
b_6}]=+36
\delta^{[a_1}_{[b_1} T^{a_2a_3]}{}_{b_2\dots b_6]} -
2.5!\delta ^{a_1a_2a_3}_{[b_1b_2 b_3}T_{b_4b_5 b_6]}
\eqno(4.9)$$
Examining the other commutators one finds that SL(32) is generated by
multiple commutators of $T^{a_1 a_2 a_3}$. 
\par 
For the effect of SL(32) on the
central charges was given in   equation (4.3). which in terms of the new
basis, becomes 
$$
[Z_{\alpha\beta}, T^{a_1\ldots a_p} ]=(\gamma^{a_1\ldots a_p}
( \gamma^a  P_a+\gamma^{a_1a_2} Z_{a_1a_2}+
\gamma^{a_1\ldots a_5}Z_{a_1\ldots a_5}))_{\alpha}{}^{\delta}
(C^{-1})_{\delta \beta}
$$
$$+
(-1)^{{p(p+1)\over 2}}  (( \gamma^a 
P_a+\gamma^{a_1a_2} Z_{a_1a_2}+
\gamma^{a_1\ldots a_5}Z_{a_1\ldots a_5})\gamma^{a_1\ldots
a_p})_{\alpha}{}^{\delta} (C^{-1})_{\delta \beta}
\eqno(4.10)$$
As a result we find that [8] 
$$
[P_b,T_{a_1a_2a_3}]=-6\eta_{b[a_1}Z_{a_2a_3]},\  
[Z^{b_1b_2},T_{a_1a_2a_3}]=-5! Z_{a_1a_2a_3}{}^{b_1b_2}+6\delta^{b_1b_2}_
{[a_1a_2}P_{a_3]}
\eqno(4.11)$$
$$[Z^{b_1\ldots b_5}, T_{a_1a_2a_3}]=- {1\over 4} 
\epsilon^{b_1\ldots b_5}{}_{c_1\ldots c_5 [a_1a_2}Z_{a_3]}{}^{c_1\ldots
c_5}+2\delta^{[b_1b_2b_3} _{a_1a_2a_3}Z^{b_4b_5]}
\eqno(4.12)$$
\par
 GL(32) first arose in the dynamics of the M theory
fivebrane which was found [8] to 
possess an unexpected  symmetry  whose generator had three
anti-symmetrised eleven dimensional space-time indices. 
This generators was shown to obey commutation relations 
that identified it with  a generator in 
contracted version of  SL(32) [8]. Furthermore, when the
dynamics of the  fivebrane was described by a non-linear realisation the
third rank  world volume gauge field strength was found to be the
Goldstone boson for a rank three generator in SL(32) which acted on the
supercharges as in equation (4.3). 
\par
It was explained in reference [8], that SL(32) is the natural
generalisation of spin(1,10), required for the point particles, to the
situation when branes are present in M theory and it acts as a brane
rotating symmetry.  Spin(1,10) can be defined is just the group  which
acts by a transformation on
$Q_\alpha$ that takes
$(\gamma^{a})_{\alpha}{}^\beta P_a$ into on object of the same form. 
However, when branes are present this last expression is replaced by the
arbitrary matrix 
$Z_{\alpha\beta}$ and so the natural symmetry to consider is SL(32). 
\par
Finally, in reference [1] it was argued that GL(32) was a symmetry of M
theory. We recall the outline of the argument. Eleven dimensional
supergravity is invariant under the supersymmetry algebra and it is also
invariant under  the eleven dimensional conformal group. The latter
follows from the formulation of eleven dimensional supergravity as a
simultaneous non-linear realisation of the conformal algebra  and
$G_{11}$ [1].  However, it has been known [10] for many years that the
only algebra that contains the conformal algebra and the supersymmetry
algebra is Osp(1/64) and so the  result follows. 
\par
This invariance is consistent with the much earlier result of reference 
[11] which   showed that eleven dimensional supergravity was invariant
under SO(16) provided one took only
$SO(1,2)\times SO(8)$ as the 
Lorentz group. This SO(16) is contained in SL(32) in a straightforward
way;  we write the 32 component spinor in terms of its 
spin(1,2)$\times$spin(8) decomposition and then the spin(16) acts on the
indices associated with the latter. 
\par
Very recently SL(32) has been considered in the context of the holonomy
of M theory [12],[13],[14]. Some of the results of these papers can be
viewed as a consquence of the SL(32) symmetry found  in [8,9,1] and
 some of the conjectures made in [13] were already
proposed in [8,9,1] and indeed shown in  [1]. 

\medskip
{\bf {5   $E_{11}$ and the identification of the central charges}}
\medskip 
It was suggested [8] that  the Lorentz algebra
should be replaced by SL(32) in the context of M theory. As such, one
might anticipate that this algebra should be contained in  $E_{11}$ and, 
in particular,  one might expect that SL(32) should be contained in the
corresponding local sub-algebra  which is the Cartan involution invariant
sub-algebra of
$E_{11}$. Such an identification  can also be seen from another
viewpoint.  The world volume gauge field strength  that occurs
in the fivebrane equation of motion is the Goldstone boson for
part of the SL(32) automorphism symmetry [9] and it is also the case that 
the  third rank gauge field of eleven dimensional supergravity is the
Goldstone boson associated with the generator $R ^{a_1a_2 a_3}$ of
$E_{11}$ [1,2]. However, these two fields appear in the world volume
dynamics in a linear combination due to their gauge symmetry and  as such
the generators of the two algebras should be related to each other. 
\par
The generators of the   Cartan involution invariant sub-algebra of
$E_{11}$ are given,  at low levels, in equations (2.17) to
(2.19) and their commutation relations in equations (2.20) and (2.21). 
Comparing these with the analogous relations for the SL(32)
automorphism algebra of  the eleven dimensional supersymmetry
algebra  given in equations    (4.8)
and (4.9) we find the same commutation relations provided  we
identify 
$$T_{a_1a_2a_3}=S_{a_1a_2a_3}, \ T_{b_1\ldots b_6}=
S_{b_1\ldots b_6} \ {\rm {and }}\ 
S_{b_1\ldots b_6}{}^{[a_1a_2,a_3]}={1\over 2}\delta_{[b_1}^{[a_3}
\epsilon^{a_1a_2]}{}_{b_1\ldots b_6] c_1\ldots c_4}
T^{c_1\ldots c_4}
\eqno(5.1)$$ 
\par
 The reason for the contraction appearing in
reference [8]  is because the  symmetry
of the fivebrane dynamics considered there  involves  the  generator
$R^{b_1b_2b_3}$ rather than
$S_{a_1a_2a_3}$.  It follows from the above  arguments that, even though
we have not introduced the supersymmetry supercharges themselves,   the  
SL(32) generators identified in
$E_{11}$ in this paper are  the ones that acts on the supercharges as in
equation (4.3). 
\par 
At first sight the rank of SL(32) is too large for it to be contained in 
$E_{11}$. However, $E_{11}$ does contain  very large commuting
sub-algebras, certainly much larger than eleven, and to identify  Sl(32)
we have set to zero some generators in $E_{11}$ and this step could also
increase the  number of  generators allowed in a commuting sub-algebra. 
The only remaining SL(32) generator that is left to identify is $T_a$ 
which occurs at the next level and it would be good to extent  the
calculation given in this paper to the next level and so identify this
generator. 
\par 
Given the above identification of SL(32)
one can examine if its action on the central charges given  in equations
(4.11) and (4.12) plays a role in
$E_{11}$. Indeed, comparing these equations with equations (3.15) and
(3.16) we find the same commutators  provided we identify the central
charges of the eleven dimensional supersymmetry algebra with the two  
and five rank anti-symmetric tensors that appear in the $E_{11}$
representation  with highest weight $l_1$, in particular we take 
$$
W^{a_1a_2}=2Z^{a_1a_2}, \ W^{a_1\ldots a_5}=2. 5! Z^{a_1\ldots a_5}, \ 
W^{a_1\ldots a_7,c}=60 \epsilon ^{a_1\ldots a_7}{}_{d_1\ldots d_4}
Z^{c d_1\ldots d_4}
\eqno(5.2)$$
Since the generator $P_a$ is in common to the supersymmetry algebra and
the semi-direct product of $E_{11}$ with its $l_1$ representation,  we 
are obliged to  identify the two 
and five rank anti-symmetric tensors that appear in 
$E_{11}\oplus_s L_1$. 
\par
The conjectured $E_{11}$ symmetry of M theory was uncovered  in reference
[2] by examining the bosonic sector  of eleven dimensional supergravity and
it is encouraging that aspects of  M theory found outside this sector,  
i.e. in the fermionic sector
and in the fivebrane dynamics lead to symmetries that are contained in
semi-direct product of $E_{11}$ with   its $l_1$ representation. This
bodes well for  the incorporation of 
supersymmetry and the rest of the conformal group into an extension of
$E_{11}\oplus_s L_1$. 
\par
The realisation that $E_{11}$ places the central charges in the same 
symmetry multiplet as the translations generators indicates that the 
theory which is invariant under $E_{11}$ should involve the
usual space-time and coordinates associated with the central charges. 
In particular, we should construct    the non-linear realisation of
$E_{11}\oplus_s L_1$ and so consider  a group element of the
form 
$$g = exp({x^a P_a+x_{a_1a_2}W^{a_1a_2}+
x_{a_1\ldots a_5}W^{a_1\ldots a_5}+\dots })
$$
$$ exp ({h_{a}{}^b K^a{}_b}) exp {({A_{c_1\dots
c_3}  R^{c_1\ldots c_3}\over 3!}+ {A_{c_1\ldots c_6}
 R^{c_1\ldots c_6}\over 6!})} \dots 
\eqno(5.3)$$
where $+\dots $ stand for higher level  generators in $l_1$ and the
final 
$\dots $ for  terms containing higher level generators of $E_{11}$. The
fields should then  depend on $x^a, x_{a_1a_2}, x_{a_1\ldots a_5},
\dots$.  
This formulation should be able to describe  point particles and 
 branes on an equal footing as it encodes the brane rotating symmetries.
Traditionally,  non-linear realisation have been used to  describe 
the low
energy dynamics of a theory in which a symmetry is spontaneously broken, 
however, at high energies the same  symmetry is expected to be linearly
realised.  It is to be expected that the same might occur here and that
the fundamental theory may have its symmetries realised on different
variables. This presumably should include the coordinates.  
\par
It would be interesting to examine the dimensional reduction of the 
above non-linear realisation and compare the result  in nine
dimensions with the T duality invariant effective action constructed in
reference [17]. 
\medskip
{\bf {6   Solutions and Weyl Transformations}}
\medskip 
In a recent paper [15], the Cartan sub-algebra of $E_{11}$, that is group
elements of the form $g=e^{q^m(x) H_m}$ where $H_m$ are the Cartan
generators of $E_{11}$, was considered and the  action of the $E_{11}$
Weyl transformations induced on the space-time fields $q^m$ was  found. 
After relating these to the diagonal components of the metric, it was
shown that the moduli space of an enlarged set of Kasner solutions carry a
representation of these Weyl transformations [15]. In this section we
generalise this work and  consider  group elements that include
the space-time translation generators and higher level generators in the 
$l_1$ representation. We will find the action of the Weyl
transformations also on the space-time  and other coordinates
at low levels. As a result, we will be able to find  the action of Weyl
transformations on the solutions themselves and not just their moduli.
Our discussion will be rather brief, but we will  set out the general
procedure and illustrate it in the context of the generalised Kasner
solutions. 
\par
Restricting the $E_{11}$ part of the group element to be in the Cartan
sub-algebra, the group element of equation (5.3) takes the form
$$g = exp({x^\mu P_\mu+x_{a_1a_2}W^{a_1a_2}+
x_{a_1\ldots a_5}W^{a_1\ldots a_5}+\ldots }) exp (q^m H_m)
$$
$$=exp({x^\mu P_\mu+x_{a_1a_2}W^{a_1a_2}+
x_{a_1\ldots a_5}W^{a_1\ldots a_5}+\ldots }) exp(p_{a}{K^a{}_a}).
\eqno(6.1)$$
For $E_{11}$ the Cartan sub-algebra is expressed in terms of the
generators ${K^a{}_a}$ which belong to GL(11) [2]. A detailed explanation
of the change from the $q^m$ to the  $p_a$ variables is given in section 2
of reference [15]. The theory resulting from the non-linear realisation
involving  such a restriction  only possess a diagonal metric whose
components are given by
$g_{\mu\mu}= e^{2p_a}\eta_{ a\mu}$.
\par
Although the Cartan sub-algebra of a  Kac-Moody algebra $g$ carries a
representation of the Weyl group this is not the case  for the 
representations of $g$. However, an extension of the Weyl group by a
cyclic group does have an action on representations of 
$g$. The effect of this extension is to introduce various signs into the
action of the Weyl group. In this first account we will omit the, possibly
significant,  effects of the centre and give the effect of Weyl
transformations are only up to signs. The Weyl reflections, 
$S_{\alpha_a} $ corresponding to the simple roots
$\alpha_a$ on the weights
$w$ of a representation are given by $S_{\alpha_a}w=w-(\alpha_a,
w)\alpha_a$. Carrying this out on the root string of equation (3.2) and
using the correspondence between  the weights and the generators, we find
that  $S_a,\ a=1,\ldots ,10$ does not introduce any $\alpha_{11}$ and so
they do  not change the level of the generator on which they act. They 
act on the space-time translation generators as 
$$S_a (P_a)=P_{a+1},\ S_a (P_{a+1})=P_{a},\ S_a (P_b)=P_{b}, b\not=a,a+1
\eqno(6.2)$$
The Weyl transformation $S_{11}$ can  change the level as it can 
introduce a $\alpha_{11}$ in its reflection. Its action on
the translations is given by 
$$S_{11} (P_a)=P_{a},\ a\not=9,10,11,
$$
$$ S_{11} (P_{9})=W^{1011},\ 
S_{11} (P_{10})=W^{911},\ S_{11} (P_{11})=W^{910},
\eqno(6.3)$$
while on the second rank central charge we find that 
$$S_{11} (W^{1011})=P_{9}, \ S_{11} (W^{911})=P_{10}, ,\ S_{11}
(W^{910})=P_{11}, \ S_{11} (W^{ab})=W^{ab},\ a\le 8,b>8 .
\eqno(6.4)$$
$ S_{11}$ takes the remaining components of the two-form central
charge into the five-form central charge. 
Examining the group element of equation (6.1) we find that 
the Weyl transformations $S_a,\ a=1,\ldots ,10$  act on the space-time
coordinates and variables $p_a$ as 
$$ x^a \ \leftrightarrow x^{a+1}, \ p_a \ \leftrightarrow p_{a+1}.
\eqno(6.5)$$
It is important to remember that the $p_a$ depend on the space-time and
other  coordinates and the change in these must also be carried out when
evaluating the transformed variable. 
The Weyl transformation $ S_{11}$ mixes the space-time coordinates with
the central charge coordinates, for example  $x^c\ \to x^c, \
c=1,\ldots , 8$ , $x^9\  \ \leftrightarrow x_{1011}$, 
  $x^{10}\  \ \leftrightarrow x_{911}$ and $x^{11}\  \ \leftrightarrow
x_{910}$. 
\par
The generalised Kasner solutions can be labeled  by the  
space-time variable that the metric depends on.  The $x^b$-Kasner
solution has a metric of the form 
$$g_{\mu\mu}=\eta_{\mu a}e^{2\tilde p_a x^b}
\eqno(6.6)$$
where $\tilde p_a$ are constants which must satisfy 
$$\sum_{c, c\not=b} \tilde p_c=\tilde p_b,\ \sum_{c, c\not=b} 
(\tilde p_c)^2=(\tilde p_b)^2
\eqno(6.7)$$
in order to obey Einstein's equations. The usual Kasner 
solution has
$b=0$ i.e. depends on a time variable. 
\par
Using equation (6.5) we can carry out the Weyl transformations for $S_a,\
a=1,\ldots ,10$ on the generalised Kasner solutions and we find that 
the solutions are indeed exchanged under the action of these Weyl
transformations.  It takes the $x^b$-Kasner solution into the
$x^b$-Kasner solution if $b\not=a,a+1$ and  it swops the  $x^a$-Kasner
solution with  the $x^{a+1}$-Kasner solution. We observe that this
includes the swopping of the temporal and spatial generalised Kasner
solutions if $a=1$. 

\medskip
{\bf {Acknowledgment}}
The author wishes to thank David Olive and Andrew Pressley 
 for discussions. He also acknowledges the support of the PPARC rolling
grant  number  PPA/G/O/2000/00451.  
\medskip
\medskip
{\bf {References}}
\medskip

\item{[1]} P.~C. West, {\it Hidden superconformal symmetry in {M}
    theory },  JHEP {\bf 08} (2000) 007, {\tt hep-th/0005270}. 
\item{[2]} P. West, {\it $E_{11}$ and M Theory}, Class. Quant.
Grav. {\bf 18 } (2001) 4443 , hep-th/0104081. 
\item{[3]} M. Gaberdiel and P. West,  {\it Kac-Moody algebras in
perturbative string theory}, JHEP {\bf  0208} (2002) 049, hep-th/0207032. 
\item{[4]} P. West, {\it Very Extended $E_8$ and $A_8$ at low levels,
Gravity and Supergravity}, Class.Quant.Grav. {\bf 20 } (2003) 2393-2406, 
 hep-th/0212291. 
\item{[5]} I. Schnakenburg and  P. West, {\it Kac-Moody 
Symmetries of IIB Supergravity,}, Phys. Lett. {\bf B517} (2001).
\item{[6]} A. Borisov and V. Ogievetsky, {\it 
Theory of dynamical affine and conformal 
symmetries as the theory of the gravitational field}, 
Teor. Mat. Fiz. {\bf 21} (1974) 329. 
\item{[7]} D. Olive, M. Gaberdiel and P. West,  {\it A Class
of Lorentzian Kac-Moody Algebras} hep-th/0205068, Nucl.Phys. 
{\bf B645} (2002)
403-437. 
\item {[8]} O. Barwald and P. West, {\it Brane Rotating symmetries and
the fivebrane equations of motion}, Phys.Lett. {\bf B476} (2000) 157-164, 
hep-th/9912226.
\item{[9]} P. West, { Automorphisms, Non-linear Realizations  and 
Branes},  JHEP {\bf 0002} (2000) 024, hep-th//0001216.  
\item{[10]} J. van Holten and A. van ~Proeyen, {\it $N=1$
supersymmetry algebras in $d=2,3,4$   mod 8}, J. Phys A, {\bf 15} (1982)
376. 
\item{[11]} B. de Wit and H. Nicolai, {\it D=11 supergravity with local
$SU(8)$ invariance}, Nucl. Phys. {\bf B274} (1986) 363; 
 B. de Wit and H. Nicolai, {\it
Hidden symmetries in D=11 supergravity}, Phys. Lett. {\bf 155B} (1985) 47;
H.  Nicolai, {\it D=11 supergravity with local $SO(16)$ invariance},
Phys. Lett. {\bf 187B} (1987) 316.
\item{[12]} M. Duff and J. Liu, {\it Hidden space-time symmetries and
generalised holonomy in M theory}, hep-th/0303140.
\item{[13]} C. Hull, {\it Holonomy and symmetry in M theory},
hep-th/0305039. The author of this reference informs the author of 
the present  paper that it will shortly be modified to take into account the  
overlap with  references [1],[8] and [9]. 
\item{[14]} G. Papadopoulos and D. Timpsis, {\it The holonomy of
the supercovariant connection and Killing spinors}, hep-th/0306117. 
\item{[15]} F. Englert, L. Houart, A. Taormina and P. West, {\it 
The Symmetries of  M theories}, hep-th/0304206. 
\item{[16]}  H. Nicolai and T.  Fischbacher, {\it Low Level
Representations for E10 and E11}, hep-th/0301017.
\item{[17]} B. de Wit and H. Nicolai, {\it Hidden symmetries, central
charges and all that}, hep-th/0011239. 

\end